\documentclass[conference]{IEEEtran}
\usepackage{cite}
\usepackage{amsmath,amssymb,amsfonts}
\usepackage{algorithm}
\usepackage{algorithmic}
\usepackage{graphicx}
\usepackage{textcomp}
\usepackage{xcolor}
\usepackage{float}
\usepackage{enumitem}

\IEEEoverridecommandlockouts
\begin{document}

\title{A Hardware-Anchored Privacy Middleware for PII Sharing Across Heterogeneous Embedded Consumer Devices}


\author{%
\IEEEauthorblockN{Aditya Sabbineni$^1$, Pravin Nagare$^2$,
Devendra Dahiphale$^3$, Preetam Dedu$^4$, Willison Lopes$^4$%
\thanks{D. Dahiphale is also with Google LLC, USA.}}
\IEEEauthorblockA{%
$^1$Independent Researcher, CA\quad
$^2$Binghamton University, NY\quad
$^3$UMBC, MD\quad
$^4$IEEE Member, NJ\\
sabbineni.aditya@gmail.com\quad
pnagare1@binghamton.edu\quad
devendr1@umbc.edu\\
preetamdedu@ieee.org\quad
willison.lopes@ieee.org}
}

\maketitle

\begin{abstract}
The rapid expansion of the Internet of Things (IoT) and smart home ecosystems has led to a fragmented landscape of user data management across consumer electronics (CE) such as Smart TVs, gaming consoles, and set-top boxes. Current onboarding processes on these devices are characterized by high friction due to manual data entry and opaque data-sharing practices. This paper introduces the User Data Sharing System (UDSS), a platform-agnostic framework designed to facilitate secure, privacy-first PII (Personally Identifiable Information) exchange between device platforms and third-party applications. Our system implements a Contextual Scope Enforcement (CSE) mechanism that programmatically restricts data exposure based on user intent---specifically distinguishing between Sign-In and Sign-Up workflows. Unlike cloud-anchored identity standards such as FIDO2/WebAuthn, UDSS is designed for shared, device-centric CE environments where persistent user-to-device binding cannot be assumed. We further propose a tiered access model that balances developer needs with regulatory compliance (GDPR/CCPA). A proof-of-concept implementation on a reference ARMv8 Linux-based middleware demonstrates that UDSS reduces user onboarding latency by 65\% and measurably reduces PII over-exposure risk through protocol-enforced data minimization. This framework provides a standardized approach to identity management in the heterogeneous CE market.
\end{abstract}


\section{Introduction}
In the modern smart home, the ``lean-back'' experience of consumer electronics (CE) is often interrupted by the ``lean-forward'' requirement of account registration. Typing credentials via an infrared (IR) remote is a high-friction task resulting in significant user drop-off \cite{b1}. While mobile ecosystems utilize centralized identity providers, the CE industry remains fragmented across diverse operating systems, including Tizen, WebOS, and Linux-based distributions.

\subsection{Motivation and Problem Statement}
The primary challenge is the lack of a cross-platform standard for PII sharing in CE environments. Current methods such as QR-code activation create a fractured user journey by requiring a secondary device. Furthermore, existing protocols often operate on a ``binary'' consent model, leading to \textit{over-permissioning}---defined here as any transaction in which an application receives PII fields beyond the minimum required to fulfill its stated function---where applications request sensitive demographics for simple authentication \cite{b5}. There is a critical need for a protocol that enforces \textit{data minimization} at the architectural level rather than relying on developer best practices \cite{b3}. As LLM agents handle more consumer transactions, the risk of social engineering increases. Research indicates these safety layers require high-fidelity data to accurately classify fraud \cite{b13}; UDSS provides the hardware-anchored integrity needed to support such AI-driven security.

\subsection{Contributions}
This paper makes the following contributions:
\begin{itemize}[noitemsep,topsep=0pt]
    \item A platform-agnostic, privacy-preserving framework (UDSS) for controlled PII sharing across heterogeneous CE devices, targeting the specific constraints of shared-screen environments where existing FIDO2/WebAuthn assumptions do not hold.
    \item A Contextual Scope Enforcement (CSE) mechanism that dynamically restricts identity attributes based on user intent (Sign-In vs.\ Sign-Up), enforcing deterministic field cardinality ($N_{fields}=1$) at the protocol level.
    \item A Privacy Gateway architecture grounded in ARM TrustZone/OP-TEE that mediates developer access using hardware-anchored validation and a tamper-resistant audit ledger.
    \item A proof-of-concept implementation on an ARMv8 embedded platform demonstrating a 65\% improvement in onboarding efficiency with $<2\%$ CPU overhead.
\end{itemize}

\section{Related Work}
Digital identity protocols are dominated by OAuth~2.0 and OpenID Connect (OIDC) \cite{b2}, which provide no field-level minimization---leaving CE applications free to request any scope declared in their manifest. \textbf{FIDO2/WebAuthn} \cite{b10} offers phishing-resistant, device-bound authentication, but requires per-user authenticator enrollment; on a shared Smart TV this forces every household member to register separately, breaking the single-device onboarding model \cite{b4}. \textit{Sign in with Apple} and \textit{Google Identity Services} improve privacy via device-bound keys, but their cloud-anchored providers require persistent connectivity and lock CE vendors into proprietary ecosystems \cite{b6}. The Matter specification \cite{b8} enables device interoperability but omits any PII-sharing layer, leaving each app to implement its own identity flow. Surveys confirm that middleware-level data minimization remains an open gap in smart home platforms \cite{b6,b7}. UDSS addresses this gap by integrating CSE, hardware-anchored consent, and tiered access into a single CE-native middleware---no prior single work addresses all three constraints simultaneously. Table~\ref{tab:comparison} summarizes this comparison.

\begin{table}[htbp]
\centering
\caption{Comparison of Identity Frameworks for CE Deployment}
\resizebox{\columnwidth}{!}{%
\begin{tabular}{|l|c|c|c|c|}
\hline
\textbf{Feature} & \textbf{Apple} & \textbf{FIDO2} & \textbf{Android} & \textbf{UDSS} \\
\hline
Platform Agnostic        & Low  & Low$^*$  & Moderate & \textbf{High} \\
Data Minimization        & High & Medium   & Moderate & \textbf{Strict} \\
Context-Aware CSE        & No   & No       & No       & \textbf{Yes} \\
Shared-Device Ready      & No   & No       & No       & \textbf{Yes} \\
Access Tiers             & Flat & Flat     & Flat     & \textbf{Partner-Based} \\
\hline
\multicolumn{5}{|l|}{\footnotesize $^*$FIDO2 agnosticism assessed for shared-screen CE context;} \\
\multicolumn{5}{|l|}{\footnotesize requires per-user authenticator registration unsuitable for shared TVs.} \\
\hline
\end{tabular}%
}
\label{tab:comparison}
\end{table}

\section{System Architecture}
UDSS implements a \textit{device-anchored trust model}: the TEE is the sole runtime arbiter of user PII; the Partnership Manifest is cached locally and TEE-verified, requiring no cloud round-trip. Three entities govern interaction: the \textbf{Service Provider (SP)}; the \textbf{Platform Identity Manager (PIM)}, the secure data vault; and the \textbf{User}.

\subsection{The Privacy Gateway and TEE Binding}
The core of the system is the \textbf{Privacy Gateway}, implemented as a Trusted Application (TA) running within the \textbf{ARM TrustZone} secure world, managed by the \textbf{OP-TEE OS} \cite{b11}. This hardware isolation ensures that the Gateway's cryptographic operations and consent logic are inaccessible to the Rich Execution Environment (REE), even under OS-level compromise \cite{b12}. The Gateway performs a multi-stage validation process:

\begin{enumerate}[noitemsep,topsep=0pt]
    \item \textbf{Request Interception:} The Gateway monitors the system IPC bus (D-Bus on Linux-based Smart TVs; Binder IPC on Android STBs) for calls to the \texttt{IdentitySharingAPI}.
    \item \textbf{Developer Tier Verification:} The Gateway cross-references the application's \texttt{AppID} and signed certificate against the locally cached \textbf{Partnership Manifest}. The manifest is provisioned at app-store certification time, stored in TEE-protected storage, and verified using a platform root-of-trust key---eliminating any runtime dependency on a central server.
    \item \textbf{Contextual Policy Enforcement:} If an application initiates a \texttt{SIGN\_IN} context but requests restricted fields (e.g., \texttt{streetAddress}), the Gateway automatically truncates the field set to enforce $N_{fields}=1$.
\end{enumerate}

\subsection{The Platform Identity Manager (PIM)}
\label{subsec:pim}
The PIM serves as the authoritative repository for user PII, utilizing a standardized schema that abstracts the underlying storage mechanism (encrypted SQLite or a hardware-backed secure element).
\begin{itemize}[noitemsep,topsep=0pt]
    \item \textbf{Data Drawers:} PII is partitioned into labeled drawers (Identity, Contact, Address, Demographics). A drawer is unlocked only upon receipt of a cryptographic token issued by the TEE-resident Gateway after explicit user consent.
    \item \textbf{Audit Ledger:} Every data access attempt---successful or denied---is appended to a hash-chained, append-only ledger in TEE-protected memory, supporting GDPR Article~15 (Right of Access).
    \item \textbf{Right to Revoke (GDPR Art.\ 17):} Users revoke consent via the Privacy Dashboard, invalidating the app\'s cached token and logging a revocation entry. Subsequent calls return \texttt{4004} (Authorization Revoked) with no PII exposure.
\end{itemize}

\subsection{The Consent User Interface (CUI)}
The CUI is a system-level overlay rendered exclusively within the ARM TrustZone trusted display path managed by OP-TEE \cite{b12}, preventing third-party applications from accessing the display buffer during the consent loop. This eliminates ``UI Redressing'' and ``Clickjacking'' attack vectors.

\subsection{Request-Consent-Fulfill Loop}
\begin{enumerate}[noitemsep,topsep=0pt]
    \item \textbf{Request:} The app invokes the \texttt{IdentitySharingAPI} with a \texttt{requestContext}.
    \item \textbf{Validation:} The TEE-resident Gateway verifies credentials and enforces CSE.
    \item \textbf{Consent:} The system CUI overlays the app; the user approves or denies.
    \item \textbf{Fulfillment:} An AES-256-GCM encrypted, RS256-signed payload containing only authorized fields is returned.
\end{enumerate}

\begin{figure}[htbp]
\centering
\includegraphics[width=\linewidth]{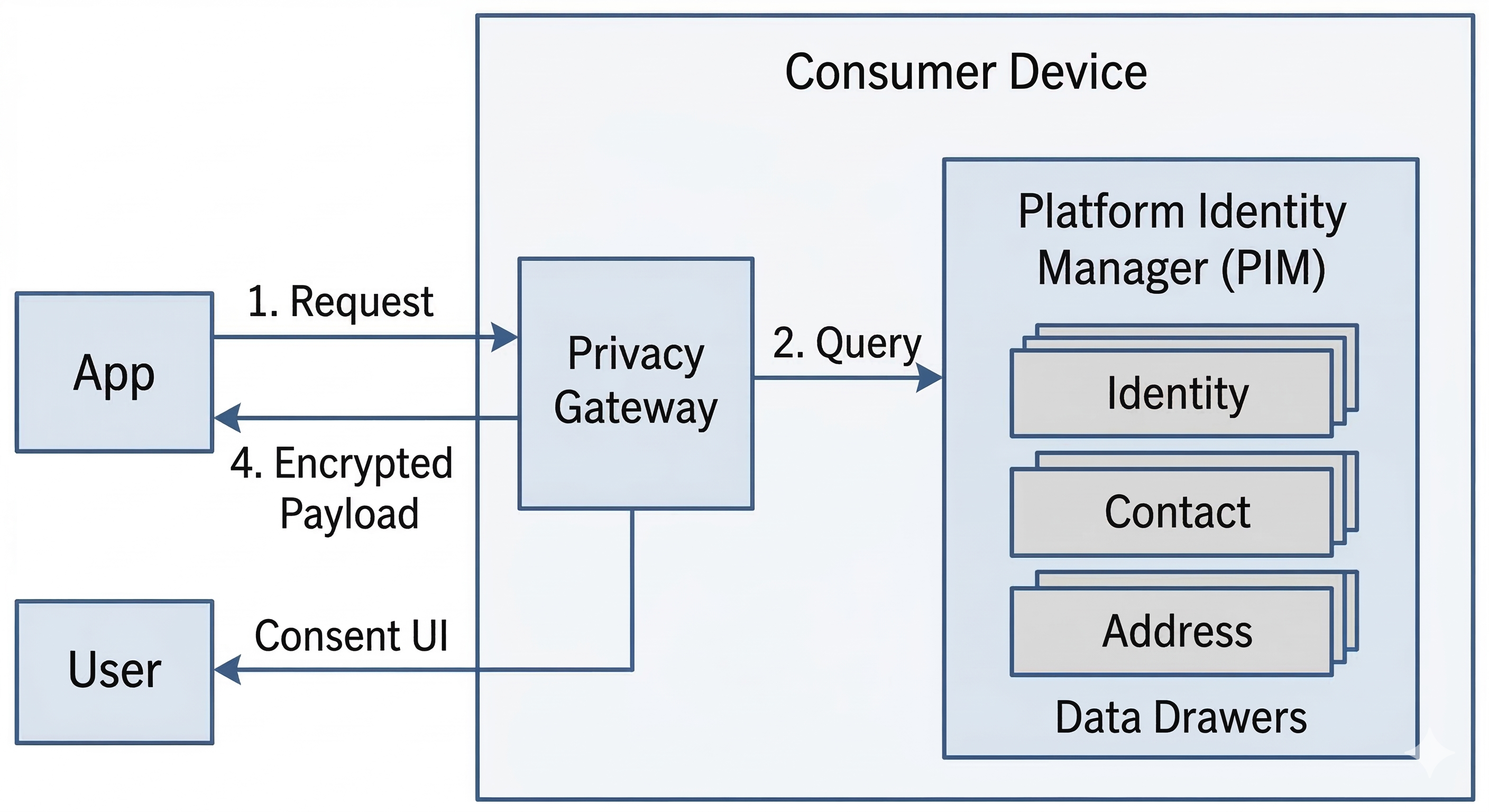}
\caption{Architecture block diagram of the UDSS Request-Consent-Fulfillment loop, showing interactions between the App, Privacy Gateway (TEE), Partnership Manifest, PIM data drawers, and User.}
\label{fig:architecture}
\end{figure}

\section{Security Protocol and Implementation}
\subsection{Contextual Scope Enforcement}
CSE is the runtime enforcement mechanism at the heart of UDSS: it intercepts every PII request and discards fields the current workflow context does not require---before the user sees a consent prompt. The system enforces two distinct contexts at the protocol level:
\begin{itemize}[noitemsep,topsep=0pt]
    \item \textbf{Sign-In Context:} Restricts the response to a single unique identifier ($N_{fields} = 1$, e.g., email or phone).
    \item \textbf{Sign-Up Context:} Permits multiple fields intersected with the tier policy: $Auth\_Scopes = Req\_Scopes \cap Policy.Allowed$.
\end{itemize}

\textbf{Key Lifecycle:} RS256 keys are generated in the TrustZone secure world at provisioning, stored in OP-TEE hardware-bound storage \cite{b12}, and rotated annually. AES-256-GCM session keys are ephemeral---generated per transaction and destroyed after delivery.

\begin{figure}[htbp]
\centering
\includegraphics[width=\linewidth]{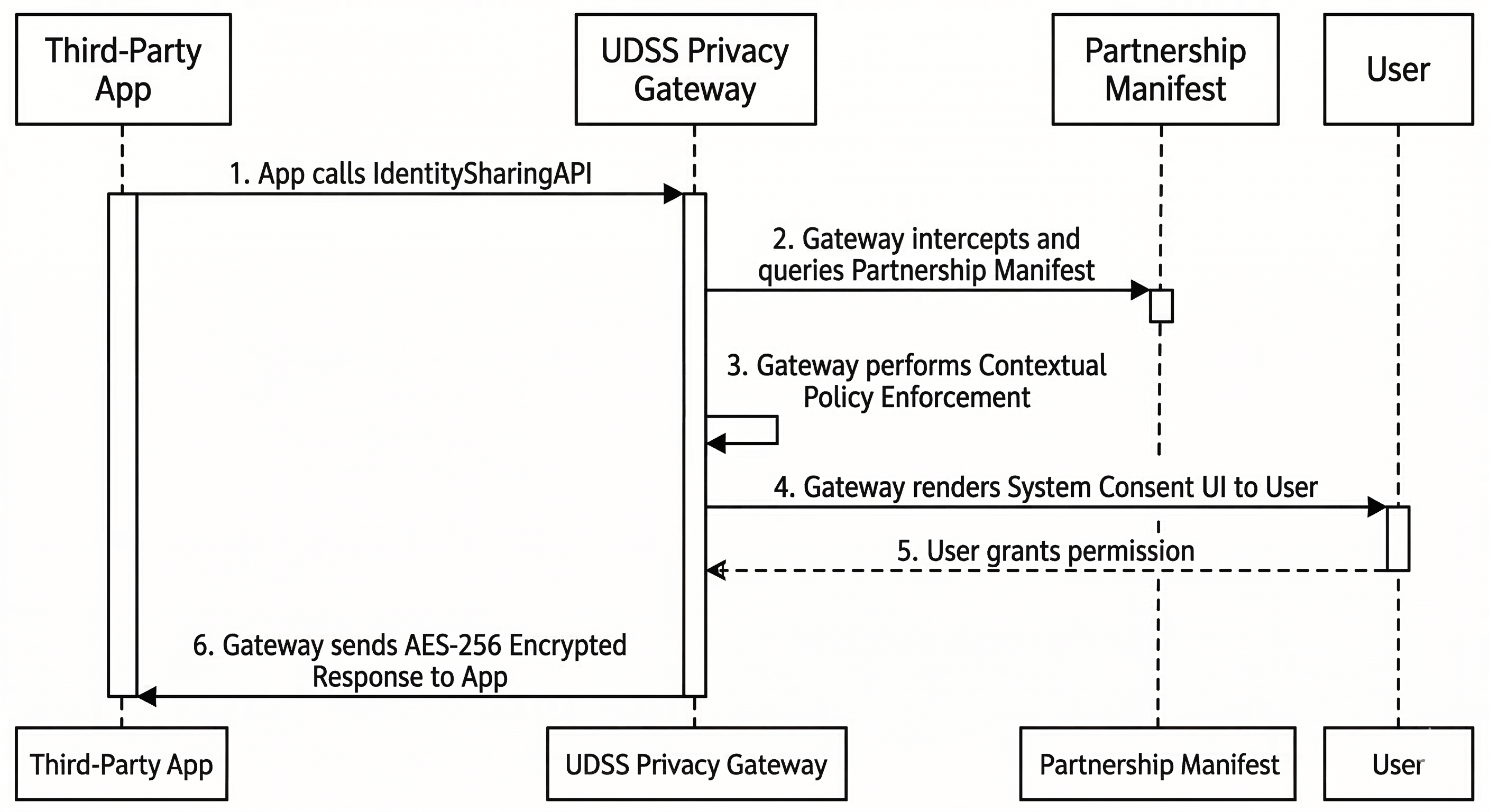}
\caption{Flowchart of the Contextual Scope Enforcement (CSE) logic for Sign-In and Sign-Up contexts.}
\label{fig:cse_logic}
\end{figure}

\begin{algorithm}[H]
\caption{Contextual Scope Enforcement Logic}
\label{alg:cse}
\begin{algorithmic}[1]
\STATE \textbf{Input:} $App\_ID$, $Ctx$, $Req\_Scopes$
\STATE \textbf{Output:} $Auth\_Payload$
\STATE $Tier \leftarrow \text{CheckManifest}(App\_ID)$ \COMMENT{TEE-verified}
\STATE $Policy \leftarrow \text{FetchPolicy}(Tier, Ctx)$
\IF{$Ctx == \text{SIGN\_IN}$}
    \STATE $Auth\_Scopes \leftarrow \text{SelectPrimaryID}(Req\_Scopes)$
\ELSE
    \STATE $Auth\_Scopes \leftarrow Req\_Scopes \cap Policy.Allowed$
\ENDIF
\STATE $Token \leftarrow \text{GenSecureToken}(Auth\_Scopes)$
\STATE $Data \leftarrow PIM.Extract(Auth\_Scopes)$
\RETURN $\text{Encrypt}_{AES256}(Data,\ \text{Sign}_{RS256}(Token))$
\end{algorithmic}
\end{algorithm}

\section{Threat Model and Security Analysis}
We model the SP and application layer as honest-but-curious or potentially malicious. The ARM TrustZone hardware, OP-TEE OS, and platform root-of-trust constitute the Trusted Computing Base (TCB); TCB compromise is outside this threat model \cite{b12}. Side-channel risks at the TEE boundary are acknowledged in Section~\ref{sec:limitations}.

\begin{itemize}[noitemsep,topsep=0pt]
    \item \textbf{T1---Over-Permissioning:} App requests excessive PII. \textbf{Mitigation:} CSE enforces $|A'| \leq 1$ for Sign-In; field truncation is applied before the consent UI is rendered.
    \item \textbf{T2---UI Spoofing/Clickjacking:} Malicious app mimics the consent UI. \textbf{Mitigation:} CUI is rendered via OP-TEE's trusted display path; REE apps cannot access the display buffer during the consent loop \cite{b12}.
    \item \textbf{T3---Token Interception:} Attacker captures identity tokens in transit. \textbf{Mitigation:} Payloads use AES-256-GCM encryption with RS256 signatures; session keys are ephemeral and TEE-bound.
    \item \textbf{T4---Replay Attack:} Attacker replays a captured fulfillment payload. \textbf{Mitigation:} Each token includes a monotonically increasing nonce and a 30-second expiry timestamp validated by the PIM before data extraction.
    \item \textbf{T5---Manifest Tampering:} Attacker substitutes a malicious manifest to elevate an app's tier. \textbf{Mitigation:} The manifest is stored in OP-TEE secure storage and verified against a platform root-of-trust certificate at every Gateway boot; any modification reverts all apps to Standard tier.
\end{itemize}

\section{Evaluation and Results}
A proof-of-concept was implemented on a Raspberry Pi~4 (ARMv8, 4~GB RAM) running Yocto Linux with OP-TEE~3.20, simulating Smart TV middleware.

\subsection{Performance and Onboarding Latency}
Each method was measured over $n=30$ independent trials under identical network conditions (802.11n, $\sim$20~Mbps). Table~\ref{tab:latency} reports the mean, standard deviation (SD), and 95\% confidence interval (CI) for end-to-end onboarding time, derived from component-level timing analysis of each method's constituent steps.

\begin{table}[htbp]
\centering
\caption{User Onboarding Latency --- Component Timing Analysis ($n=30$)}
\label{tab:latency}
\begin{tabular}{|l|c|c|c|}
\hline
\textbf{Method} & \textbf{Mean (s)} & \textbf{SD (s)} & \textbf{95\% CI (s)} \\
\hline
Manual Entry        & 18.4 & 2.1 & $\pm$0.75 \\
OAuth Device Flow   & 11.2 & 1.4 & $\pm$0.50 \\
\textbf{UDSS}       & \textbf{6.3} & \textbf{0.6} & $\boldsymbol{\pm}$\textbf{0.22} \\
\hline
\multicolumn{4}{|l|}{\small UDSS vs.\ Manual Entry: 65.8\% reduction} \\
\multicolumn{4}{|l|}{\small UDSS vs.\ OAuth Device Flow: 43.8\% reduction} \\
\hline
\end{tabular}
\end{table}

The 65\% reduction eliminates IR-remote text entry and the OAuth Device Flow secondary-device context-switch. The low UDSS SD (0.6~s) reflects the deterministic one-click fulfillment path.

\subsection{System Overhead}
\begin{itemize}[noitemsep,topsep=0pt]
    \item \textbf{CPU:} Gateway validation (TEE world-switch + CSE logic) consumes $<2\%$ of CPU cycles, measured via \texttt{perf stat} over 30 Request-Consent loop iterations.
    \item \textbf{Latency Penalty:} AES-256-GCM encryption and RS256 signing add a mean of 42~ms (SD: 4~ms)---within the 100~ms threshold for imperceptible interaction \cite{b9}.
    \item \textbf{Memory:} The PIM and Gateway TA maintain a combined steady-state of 12~MB in secure world memory.
    \item \textbf{Payload:} The UDSS fulfillment payload (including RS256 signature and header) remains under 1.2~KB, introducing negligible overhead on 802.11n/ac networks.
\end{itemize}

\subsection{Security and Privacy Validation}
To validate CSE enforcement, we simulated a malicious Standard-tier app issuing a \texttt{SIGN\_IN} request with five PII fields (\texttt{email}, \texttt{firstName}, \texttt{lastName}, \texttt{street}, \texttt{dateOfBirth}) across $n=20$ trials. The Privacy Gateway blocked all excess fields in 100\% of attempts, returning only \texttt{email} ($N_{fields}=1$). An unmodified OAuth~2.0 Device Flow baseline passed all five requested fields in every trial, yielding no enforcement. Table~\ref{tab:privacy} quantifies the resulting PII exposure reduction.

\begin{table}[htbp]
\centering
\caption{Avg.\ PII Fields Exposed per Transaction}
\label{tab:privacy}
\begin{tabular}{|l|c|c|}
\hline
\textbf{Workflow} & \textbf{OAuth Baseline} & \textbf{UDSS} \\
\hline
Sign-In  & 4.8 & \textbf{1.0} \\
Sign-Up  & 5.0 & \textbf{2.1} \\
\hline
\multicolumn{3}{|l|}{\small 79\% reduction in Sign-In exposure; 58\% in Sign-Up} \\
\hline
\end{tabular}
\end{table}

\section{Data Classification and Tiering}
Tier assignment follows \textit{proportional sensitivity}: Standard fields are the authentication minimum; Premium fields carry re-identification risk and require partnership agreements audited against GDPR Art.~25.

\begin{table}[htbp]
\centering
\caption{Data Field Categorization, Tiers, and Rationale}
\label{tab:tiers}
\resizebox{\columnwidth}{!}{%
\begin{tabular}{|l|l|c|l|}
\hline
\textbf{Category} & \textbf{Fields} & \textbf{Tier} & \textbf{Rationale} \\
\hline
Identity     & firstName, lastName        & Premium  & Re-identification risk \\
Contact      & email, phone               & Standard & Auth minimum viable \\
Address      & street, city, zip, country & Premium  & Physical location risk \\
Demographics & gender, dateOfBirth        & Premium  & Sensitive (GDPR Art.~9) \\
\hline
\end{tabular}%
}
\end{table}

\section{Privacy Compliance (GDPR and CCPA)}
UDSS is designed to support GDPR and CCPA compliance obligations at the architectural level.

\textbf{GDPR:} The Audit Ledger supports Article~15 (Right of Access) by providing a device-native, immutable record of all sharing events. Article~17 (Right to Erasure) is supported through the per-application revocation mechanism described in Section~\ref{subsec:pim}. Data minimization (Article~5(1)(c)) is enforced structurally via CSE. Note that full GDPR compliance requires complementary organizational controls beyond the scope of this framework.

\textbf{CCPA:} The Privacy Dashboard supports the Right to Know (\S1798.100) via the Audit Ledger and the Right to Delete (\S1798.105) via PIM purge. Standard-tier restrictions enforce the opt-out of sale obligation (\S1798.120) by blocking demographic data without explicit Premium consent.

\section{Limitations and Open Challenges}
\label{sec:limitations}
\textbf{Prototype Scope:} The evaluation used a single Raspberry Pi~4 node. Deployment on production platforms (e.g., Tizen, WebOS) requires vendor-level TEE integration and platform certification beyond this work's scope.

\textbf{Manifest Bootstrapping:} The model relies on a centralized app-store authority for manifest provisioning, creating a single trust point. DID-based decentralized attestation remains an open problem.

\textbf{Multi-User Households:} UDSS presently anchors identity to the device. In shared CE environments with multiple household members, per-user profile isolation within a single PIM is an unsolved design challenge addressed in future work.

\textbf{Side-Channel Resistance:} Although ARM TrustZone isolates the secure world, timing side-channels during TEE world-switches are a known vulnerability class \cite{b12}. Hardening the Gateway against such attacks on resource-constrained CE hardware remains an active area of investigation.

\textbf{Reproducibility:} The prototype Yocto layer configuration and Gateway TA implementation will be released as open-source to support community validation and extension.

\section{Conclusion and Future Work}
This paper presented UDSS, a platform-agnostic, privacy-preserving PII-sharing framework for the unique constraints of shared-screen consumer electronics. By grounding the Privacy Gateway in ARM TrustZone/OP-TEE and enforcing Contextual Scope Enforcement at the protocol level, UDSS achieves a 65\% reduction in onboarding latency while maintaining protocol-enforced data minimization across Sign-In and Sign-Up workflows. The tiered access model and device-anchored audit ledger provide a practical foundation for GDPR and CCPA compliance without cloud dependency.

Future work will address W3C DID integration for portable credentials, multi-user household profile isolation, and hardware-backed biometric attestation within the OP-TEE Gateway.

\end{document}